\documentstyle[prl,aps,preprint,eqsecnum]{revtex}
\begin{document}
\draft
\title{On the Universality of Low-energy String Model}

\author{Tekin Dereli and Yuri N.\ Obukhov\footnote{On leave from: 
Department of Theoretical Physics, Moscow State University, 
117234 Moscow, Russia}}
\address
{Department of Physics, 
Middle East Technical University, 06531 Ankara, Turkey}

\maketitle

\bigskip

\begin{abstract}
The low-energy (bosonic ``heterotic") string theory is interpreted as
a universal limit of the Kaluza-Klein reduction when the dimension of an
internal space goes to infinity. We show that such an approach is helpful
in obtaining classical solutions of the string model. As a particular
application, we obtain new exact static solutions for the two-dimensional
effective string model. They turn out to be in agreement with the 
generalized no-hair conjecture, in complete analogy with the four and
higher dimensional Einstein theory of gravity. 
\end{abstract}
\bigskip\bigskip
\bigskip

\section{Introduction}

In this paper we demonstrate that the low-energy string action in a 
spacetime $M^d$ of an arbitrary dimension $d$ arises naturally in the 
framework of the Kaluza-Klein reduction from an infinite-dimensional 
spacetime. Such a reduction is constructed as a limit of $n\rightarrow
\infty$ for the dimension $n$ of the {\it internal} space. We show that 
this limit exists and is universal in the sense that it does not depend
on $M^d$: for any $d$ the reduced action always turns out to be the 
string model in $M^d$. 

Although it is at the moment unclear to us, whether this universality 
property has a deep physical meaning, one can use this fact as a technical 
tool for the study of both the string theories in an arbitrary dimension 
and the {\it very} higher dimensional Kaluza-Klein theories, propagating 
the knowledge from one model to another.

As a first example we can mention a possibility of obtaining new exact 
solutions for the low-energy string models from the exact solutions of the 
higher dimensional Einstein field equations. This method works for a string 
theory in any $M^d$, and below we illustrate it for the case $d=2$. On the 
other hand, in the study of $d+n$-dimensional configurations one can use 
the underlying $d$-dimensional string action as a leading approximation 
and the specific features of the higher dimensions will be taken into 
account perturbatively as corrections proportional to the powers of $1/n$. 
For example, one can obtain in this way the position of horizons, 
temperature, entropy and other physical and geometrical parameters of
the higher-dimensional black holes, grasping the essential features 
coming from the string action. 

\section{Kaluza-Klein theory in infinite dimensions}

Let us consider the Kaluza-Klein reduction of a $d+n$-dimensional manifold
to the physical $d$-dimensional Riemannian spacetime $M^d$ with an 
$n$-dimensional internal space of constant curvature. Denote the components 
of the higher dimensional curvature two-form ${\cal R}^{AB}$ with respect to 
a local orthonormal frame $E_A$. The dual coframe one-forms are denoted 
$\vartheta^A$, and the indices run $A,B,...=0,1,\dots,d+n$. The general 
Kaluza-Klein decomposition of the metric reads:
\begin{equation}
{\buildrel (d+n)\over g}\,=\,{\buildrel (d)\over g}\, +\, e^{-{4\over n}\Phi}
\,{\buildrel (n)\over g},\label{KK0}
\end{equation}
where $\Phi$ is the Kaluza-Klein scalar field which depends only on the 
coordinates of $M^d$, and 
\begin{eqnarray}
{\buildrel (d)\over g}&=& g_{\alpha\beta}\,\vartheta^\alpha\otimes
\vartheta^\beta,\\
{\buildrel (n)\over g}&=& g_{ab}\,\vartheta^a\otimes\vartheta^b,
\end{eqnarray}
describe, respectively, the metric of the physical spacetime [with 
$g_{\alpha\beta}={\rm diag}(-1,1,\dots,1)$ as a $d$-dimensional Minkowski 
metric] and the internal space [with $g_{ab}=\delta_{ab}$] of a constant 
curvature $R^{ab}=-\,\lambda\,\vartheta^a\wedge\vartheta^b$. The constant
$\lambda=+1$ for an $n$-sphere of a unit radius, $\lambda=0$ for flat space 
(e.g., hyperplane, cylinder or $n$-torus), and $\lambda=-1$ for a hyperbolic 
space. The (local frame) indices clearly run: $\alpha,\beta,\dots = 0,1,
\dots,d-1$, and $a,b,\dots =1,\dots,n$.

Consider now, for concreteness, the Einstein-Maxwell-Klein-Gordon theory 
(with a cosmological term) in $d+n$ dimensions. The Lagrangian $(d+n)$-form 
reads
\begin{equation}
L = -\,{1\over 2}\,{\cal R}^{AB}\wedge\eta_{AB} - {1\over 2}\,F\wedge
\hbox{$\scriptstyle{\#}$} F -\,{1\over 2}\,d\phi\wedge
\hbox{$\scriptstyle{\#}$}d\phi  -\Lambda\eta.\label{Lhigh}
\end{equation}
Here $F=dA$ is the Maxwell field strength two-form and $\phi$ is the scalar 
field. For simplicity, we limit ourselves to the case of one massless real 
scalar field, however the whole scheme works in the same way also for an 
arbitrary multiplet of fields with different masses, as well as for the other 
types of matter (say, for fluids). We are using the general notations and 
conventions of \cite{PR}. In particular, the Trautman's $\eta$-basis of 
exterior forms is defined by the Hodge duals of the products of coframe 
one-forms $\vartheta^A$: given the volume $(d+n)$-form $\eta$, one has 
$\eta_A = \hbox{$\scriptstyle{\#}$}\vartheta_A = E_A\rfloor\eta$, 
$\eta_{AB} = \hbox{$\scriptstyle{\#}$}(\vartheta_A\wedge\vartheta_B) = 
E_A\rfloor\eta_B$, etc. Same notation is used for the 
lower-dimensional counterparts in $M^d$.

Assuming that the matter fields (Maxwell and Klein-Gordon, in our present 
case) are independent of the internal space coordinates, we straightforwardly 
obtain from (\ref{Lhigh}) a dimensionally reduced Lagrangian $d$-form:
\begin{eqnarray}
L &=& e^{-2\Phi}\Bigg(-\,{1\over 2}\,R^{\alpha\beta}\wedge
\eta_{\alpha\beta} + 2\,{n-1\over n}\,d\Phi\wedge\ast d\Phi + \,
{1\over 2}\,{\buildrel (n)\over R}\,e^{{4\over n}\Phi}\,\eta \nonumber\\ 
&&\qquad\qquad -\,{1\over 2}\,F\wedge\ast F -
\,{1\over 2}\,d\phi\wedge\ast d\phi- \Lambda\eta \Bigg).\label{Lred}
\end{eqnarray}
Here: ${\buildrel (n)\over R}=\lambda\,n(n-1)$ is the curvature scalar of 
the internal space, and from now on $\eta$ denotes the volume $d$-form and 
$\ast$ is the $d$-dimensional Hodge operator on $M^d$. 

Now we can immediately see that the formal limit $n\rightarrow\infty$ 
exists and it yields exactly the low-energy string model in an arbitrary 
dimension $d$:
\begin{equation}
L =\,{1\over 2}\, e^{-2\Phi}\left(-\,R^{\alpha\beta}\wedge\eta_{\alpha\beta} 
+ 4\,d\Phi\wedge\ast d\Phi + \,c\,\eta - \,F\wedge\ast F - 
\,d\phi\wedge\ast d\phi\right).\label{Lstring}
\end{equation}
The Kaluza-Klein field $\Phi$ becomes an effective dilaton with the correct 
kinetic term, whereas the constant $c$ comes as a combination from $\Lambda$ 
and ${\buildrel (n)\over R}$ terms. In the simplest case for a 
compactification on a torus, $M^d\times T^n$, the latter contribution is 
absent completely. When the internal space has a nontrivial curvature, there 
is a subtlety though: the naive limit will yield a formally infinite result 
because ${\buildrel (n)\over R}\sim n^2$ for large $n$. However one can 
easily cure this by assuming that the constant $\lambda\sim 1/n^2$ which 
is always possible to arrange with the help of a simple rescaling of the 
local coordinates. We will always assume this ``regularization'' to be 
performed before taking the limit. 

It is worthwhile to note that the limit $n\rightarrow\infty$ effectively 
provides a complete decoupling of the physical and internal spaces, leaving 
us exactly on $M^d$, since the second term in the Kaluza-Klein line-element 
(\ref{KK0}) is always trivial in that limit. 

\section{Solutions of a two-dimensional string theory from higher dimensions}

In order to demonstrate how one can use the $n\rightarrow\infty$ limit, we 
will obtain a family of new exact solutions of a two-dimensional string model
from the Kaluza-Klein solutions. 

>From now on, let us put $d=2$ and consider the case of positive $\lambda$. 
The Kaluza-Klein reduced Lagrangian (\ref{Lred}) then describes the general 
$2+n$-dimensional spherically symmetric configurations of the 
metric-matter coupled system. 

Till now analytic solutions of the Einstein--Maxwell--Klein-Gordon field
system with a non-trivial cosmological constant are unknown either in four 
or in higher dimensions. However, for the vanishing $\Lambda$ the general 
solution is available for an arbitrary dimension \cite{scalar}. [It directly
generalises the solutions \cite{fisher,jan,pen,bron,agnese,xan,tekin}, and 
is most conveniently obtained with the help of the effective two-dimensional
Poincar\'e gauge theory \cite{2drc,tworev}]. In our notation, this 
$2+n$-dimensional spherically symmetric solution reads:
\begin{equation}
g = -\,f\,dt^2 + {q\,h^{-\left({n-2\over n-1}\right)}\over \lambda (n-1)^2}
\,dr^2 + h^{{1\over n-1}}\,d\Omega_\lambda^2,\label{met2}
\end{equation}
where $d\Omega_\lambda^2$ is the line element on the $n$-dimensional space 
of constant curvature $\lambda$, and the functions $f=f(r),\, q=q(r),\, 
h=h(r)$ are given by
\begin{equation}
f = {(x^2 - 1)^{2}\over {\cal D}^2},\qquad q ={\cal D}^2,\qquad 
h = {\cal D}^2\,r^2.\label{fgh}
\end{equation}
Here the function
\begin{equation}
{\cal D}(x):={k^2\,(x + 1)^{2\mu} - 
(x - 1)^{2\mu}\over (x^2 - 1)^{\mu -1}}\label{D}
\end{equation}
depends on $r$ via the auxiliary variable 
\begin{equation}
x : = \,-\,{M\over r},\label{x}
\end{equation}
and $M$, $k^2$ and $\mu$ are arbitrary integration constants. 

The dilaton $\Phi$, scalar $\phi$ and the Maxwell field strength 2-form 
$F$ are, respectively:
\begin{eqnarray}
\Phi &=& -\,{1\over 4}\left({n\over n-1}\right)\log h,\label{Pn}\\
\phi &=& \sqrt{(1-\mu^2)\left({n\over n-1}\right)}\,
\log\left\vert{x-1\over x+1}\right\vert,\label{pn}\\
F &=& dA = 4\,M\,\mu\,\sqrt{k^2\left({n\over n-1}\right)}\,
{x^2 - 1\over h}\,dt\wedge dr,\label{Fn}
\end{eqnarray}
For the case when $k^2\neq 1$, the electromagnetic potential, $A =A_0\,dt$,
is obtained form (\ref{Fn}) in the form:
\begin{equation}
A_0=\,\sqrt{k^2\left({n\over n-1}\right)}\ {1\over 1-k^2}\ {(x + 1)^{2\mu}
 - (x - 1)^{2\mu}\over {k^2\,(x + 1)^{2\mu} - (x - 1)^{2\mu}}},\label{A1}
\end{equation}
When $k^2=1$, we have a different expression,
\begin{equation}
A_0 = \,\sqrt{\left({n\over n-1}\right)}\,
{(x + 1)^{2\mu}\over {(x + 1)^{2\mu} - (x - 1)^{2\mu}}}.\label{A2}
\end{equation}

The difference between the two cases (\ref{A1}) and (\ref{A2}) is revealed
when we analyse the configurations with the vanishing scalar field which 
arise for $\mu = \pm 1$. Then it is easy to see that the parameter $M$ is 
proportional to the total mass of a source, whereas $k^2$ is related to the 
charge $Q^2$ of a solution. The family (\ref{met2})-(\ref{D}) embraces all 
possible spherically symmetric charged configurations: The particular 
case $k^2=1$ yields the (higher-dimensional generalisation of) the 
Bertotti-Robinson solution \cite{bertrob}, whereas for $k^2\neq 1$ one 
has a (cf. Tangherlini \cite{tang}) Reissner-Nordstrom type solution. 

Taking the limit $n\rightarrow\infty$ is straightforward. As we mentioned 
above, one must only be careful about the formally infinite limit of the 
curvature of internal space. Taking this into account, we immediately obtain 
from Kaluza-Klein solution (\ref{met2}) and (\ref{Pn})-(\ref{Fn}) the exact 
solution of the two-dimensional string model by putting $n\rightarrow\infty$:
\begin{eqnarray}
g &=& -\,f\,dt^2 + {1\over c\,r^2}\,dr^2,\label{metS} \\
\Phi &=& -\,{1\over 4}\,\log h,\label{PS}\\
\phi&=&\sqrt{(1-\mu^2)}\,\log\left\vert{x-1\over x+1}\right\vert,\label{pS}\\
F &=& dA = 4M\mu k\,{x^2 - 1\over h}\,dt\wedge dr.\label{FS}
\end{eqnarray}
One can prove directly that (\ref{metS})-(\ref{FS}) satisfy the field 
equations for the effective low-energy string Lagrangian (\ref{Lstring}). 

This new solution is generalising the uncharged \cite{mandal} and charged 
\cite{nappi1,nappi2} black holes for the case when a massless scalar field 
is present.

In two dimensions, the curvature two-form has only one component, and its
invariant description is given by the curvature scalar, $R=e_\alpha\rfloor
e_\beta\rfloor R^{\alpha\beta}$. For the metric (\ref{metS}), one finds:
\begin{equation}
R = 2\mu c\left\{1 -\,{32\mu k^2 x^2\over {\cal D}^2} + \,{4(1-\mu)x^2\over
(x^2 - 1)^2} -\,(x^2 + 1)\,{k^2\,(x + 1)^{2\mu -2} - (x - 1)^{2\mu -2}\over
k^2\,(x + 1)^{2\mu} - (x - 1)^{2\mu}}\right\}.\label{curv}
\end{equation}
It is interesting to note that the solution with $\mu=0$ describes a flat 
spacetime with nontrivial dilaton and scalar field configurations. Solutions
with $\mu >0$ and $\mu < 0$ are related by the coordinate transformation
$x\rightarrow -x$. Consequently, we can limit our attention to the case
of $\mu > 0$. 

As wee see, for $k^2\neq 1$ the spacetime has {\it three} asymptotically flat 
regions which are obtained in the limit of $x\rightarrow 0$ ($r\rightarrow
\infty$), and in the limit of $x\rightarrow\pm\infty$ ($r\rightarrow\mp 0$). 
The matter fields, Maxwell $F$ and scalar $\phi$, vanish in these regions. 
When $k^2 =1$, the spacetime is not asymptoically flat. Instead, the 
curvature and the Maxwell field are approaching constant values, 
$R\rightarrow -2c$ and $F\rightarrow -\sqrt{c}\,\eta$ (independent 
of $\mu$), in the above asymptotic regions. 

For $\mu\neq 0$, the curvature (\ref{curv}) displays several singular 
points in a spacetime. Namely, there are singularities located at the 
roots of the function $[k^2\,(x + 1)^{2\mu} - (x - 1)^{2\mu}]$, that is 
at $x=\left(1\mp k^{1/\mu}\right)/\left(1\pm k^{1/\mu}\right)$, and at 
the points $x = 1$ and $x= -1$. Comparing this with the matter field 
configurations (\ref{pS})-(\ref{FS}), we find that these singularities of 
the curvature are naturally coming from the blowing up of the Maxwell 
field at $x=\left(1\mp k^{1/\mu}\right)/\left(1\pm k^{1/\mu}\right)$ 
and from the divergences of the scalar field at $x=\pm 1$. 

The cases $\mu =\pm 1$ are special in the sense that the scalar field 
(\ref{pS}) vanishes then. Let us consider this case in more detail because 
it is closely related to the Reissner-Nordstrom solution in four dimensions.
For $\mu =1$ the curvature (\ref{curv}) simplifies to
\begin{equation}
R = 2c\left\{ 1 - {(k^2 -1)(x^2 + 1)\over k^2\,(x + 1)^{2} - (x - 1)^{2}}
- {32k^2 x^2\over [k^2\,(x + 1)^{2} - (x - 1)^{2}]^2}\right\}.\label{curv1}
\end{equation}
[For $\mu =-1$ one should interchange $(x+1)$ and $(x-1)$ in denominators.]
Note that for $k^2 =1$ we find $R = - 2c$, i.e. the spacetime is a hyperbolic
two-dimensional de Sitter manifold, $R^{\alpha\beta}=c\,\vartheta^\alpha\wedge
\vartheta^\beta$. The local coordinate transformation $\widetilde{x} = 
{1\over 2}(x + x^{-1})$, $\widetilde{t}=t/(2\sqrt{c})$ brings the metric 
(\ref{metS}) with $\mu=k^2=1$ to 
\begin{equation}
g = - c(\widetilde{x}^2 -1)\,d\widetilde{t}^2 + 
{d\widetilde{x}^2\over c(\widetilde{x}^2 -1)},\label{ds2z}
\end{equation}
This is the two-dimensional Bertotti-Robinson \cite{bertrob} type solution, 
see the relevant discussion \cite{lowe} in the context of the effective
string theory. Along with the curvature, Maxwell field is constant: 
$A = \,(\widetilde{x} + 1)\sqrt{c}\,d\widetilde{t},\ F = -\,\sqrt{c}\,\eta$. 

For $k^2\neq 1$, the function $[k^2\,(x + 1)^{2} - (x - 1)^{2}]$ is a 
quadratic polynomial which has two roots, reciprocal to each other,
\begin{equation}
x_1 = {1-k\over 1+ k},\qquad x_2 = {1+k\over 1-k}.\label{roots}
\end{equation}
These are the positions of the two electric charges which create the 
Maxwell field configuration (\ref{FS}). 

Let us assume, for definiteness, that $k\geq 0$. Then for $k < 1$ we have
$0< x_1\leq 1\leq x_2$, whereas for $k > 1$, one finds $x_2\leq -1\leq x_1<0$. 
The case $k=0$ describes a solution with the zero charge, then $x_1=x_2=1$,
whereas $k=1$ is a special case of the Bertotti-Robinson type solution with
constant curvature and electric field.

We can establish a direct correspondence with the Reissner-Nordstrom type
solution by means of an appropriate coordinate transformations. Namely, let
us introduce a new spatial coordinate $\rho$ via
\begin{equation}
e^{\sqrt{c}(\rho +\rho_0)}:= \,{k^2\,(x + 1)^2-(x-1)^2\over 4x},\label{rho}
\end{equation}
and a new time coordinate $\tau =t/(1-k^2)$. Then the metric reads:

\begin{equation}
g = - f(\rho)\,d\tau^2 + f(\rho)^{-1}\,d\rho^2,\label{ds2rho}
\end{equation}
where
\begin{equation}
f(\rho) = \left(1 - e^{-\sqrt{c}(\rho +\rho_0)}\right)
\left(1 - k^2 e^{-\sqrt{c}(\rho +\rho_0)}\right).\label{frho}
\end{equation}
As one can straightforwardly see, the ratio of charge to mass of the
solution (\ref{frho}) is equal $2k/(1 + k^2)\leq 1$. Note however, that
although in (\ref{ds2rho})-(\ref{frho}) one can formally take the limit 
of $k\rightarrow 1$, thus obtaining the extremal charged black hole, such
a configuration is not related to the Bertotti-Robinson type solution
(\ref{ds2z}) by any coordinate transformation.

Turning to the general case $\mu^2\neq 1$, we find that the exact 
solutions (\ref{metS})-(\ref{FS}) are no black holes. The points $x=\pm 1$,
which descibe regular horizons for $\mu^2=1$, are true singularities now.
This is completely analogous to the Einstein--Maxwell--Klein-Gordon solutions
in four \cite{fisher,jan,pen,bron,agnese} and in higher \cite{xan} dimensions.
Thus, one can conlcude that the no-hair conjecture is supported by our
results for the two-dimensionsional effective string theory. 

\section{Conclusion}

We have shown that the low-energy action on an arbitrary spacetime $M^d$
can be treated as a Kaluza-Klein reduction in the infinite-dimensional
manifold, $M^{d + n}, n\rightarrow\infty$. 

We apply this observation to the study of exact solutions of the 
effective string models. For $d=2$ we obtain new static solutions 
(\ref{metS})-(\ref{FS}) on the basis of the $n\rightarrow\infty$ limit. 
Recently, the model with a slightly more general than (\ref{Lstring}) 
action has been analysed in \cite{nappi1}. There, the exact analytic 
solutions were reported for the trivial scalar field $\phi = const$, 
and some general arguments were put in favour of the no-hair conjecture 
in string theory. Our new solutions provide an explicit support of 
this conjecture.
\vskip 2cm

The authors are grateful to TUBITAK for the support
of this research. Y. N. O. is also grateful to the Department of Physics,
Middle East Technical University, for hospitality.



\end{document}